\magnification=1200
\pretolerance=10000
\baselineskip=24 pt
\centerline {\bf Glitches, torque evolution and the dynamics}
\centerline {\bf of young pulsars}
\vskip 1 true cm
\centerline {M.P.Allen and J.E.Horvath}

\vskip 1 true cm
\centerline {\it Instituto Astron\^omico e Geof\'\i sico}
\centerline {\it Universidade de S\~ao Paulo - Av. M.St\'efano 4200 - 
\'Agua Funda}
\centerline {\it (04301-904) S\~ao Paulo SP - Brasil}
\centerline {\it foton@andromeda.iagusp.usp.br}
\centerline {\it mpallen@orion.iagusp.usp.br}

\vskip 1 true cm
\noindent
{\bf ABSTRACT}

The Crab pulsar has suffered in 1975 and 1989 two glitches in which the 
frequency did not relaxed to the extrapolated pre-glitch value but 
rather spun up showing long-term changes in the frequency derivative 
$\dot \Omega$. This particular behaviour has been interpreted as evidence for
an evolution of the torque acting upon the star. A variable torque may be
related to non-canonical braking indexes, for which some determinations have
been possible. We briefly analyse in this work the consistency of 
postulating a growth in the angle between the magnetic moment and the 
rotation axis as the cause of such events. 
We show that this hypothesis leads to the determination of the initial period,
initial and present angles, according to the assumed angle growth, for young 
pulsars whose respective braking indices $n_{obs}$ and jerk parameters 
$m_{obs}$ are known, and some insights on the equation of state.
\vskip 1 true cm
\noindent
\vfill\eject
\noindent
{\bf 1 INTRODUCTION}

The dynamics of compact stars offers a unique opportunity to explore and 
understand the physics of dense matter. Particularly, timing irregularities 
have proved to be extremely important (and challenging) for building a 
coherent picture of those stellar interiors involving hadronic matter at 
subnuclear and supranuclear densities.

Among the most notorious timing irregularities, sudden discontinuities 
({\it glitches}) are observed in the pulsar frequency $\Omega \; and \;$ 
spin-down rate $\dot \Omega$. Several of such events have been observed from 
the Crab, Vela and some other pulsars. It is generally agreed that they are 
the result of a (complex) interplay between the superfluid component(s) and 
the rest of the star. Theoretical models searched an explanation in terms of 
starquakes (Baym et al.1969) or, more recently, vortex motion in the 
superfluid component (Alpar et al. 1984, Pines \& Alpar 1985, Link, Epstein \& 
Baym 1993). While the latter model is flexible enough to accomodate a large 
body of observations, evidence coming from very timely and detailed 
observations of two glitches in the Crab pulsar seems to indicate that new 
physical inputs may be required to explain the data (see below). Since the 
Crab can be considered as the best studied object, it may be presumed that an 
analogous behaviour could be present in the young pulsar sample for which an 
increasing body of data is available. We shall attempt to understand the 
dynamics of the youngest objects and build a consistent picture of their 
evolution.

In 1975 and 1989 the spin rate of the Crab pulsar $\Omega$ suddenly increased 
by amounts $\Delta\Omega / \Omega \; \sim \;10^{-8}$ and after that continued 
to spin-down at a faster rate (that is, the pulsar continued to spin {\it 
slower} than before) of $\Delta \dot \Omega / \dot \Omega \; \sim \; 10^{-4}$ 
(Gullahorn et al. 1977, Lohsen 1981, Lyne, Smith \& Pritchard 1992). The same 
feature is also present in the 1969, 1981 and 1986 events (Lyne \& Pritchard 
1987, Lyne, Pritchard \& Smith 1993). This behaviour has been attributed to a 
decoupling of some internal shell or a change in the external torque acting on 
the star (Gullahorn el al. 1977, Demia\'nski \& Pr\'oszy\'nski 1983). 
By using a general postglitch relaxation equation of the form 

$$ {\partial \omega(r,t) \over {\partial t}} \; = \; f(\omega,r) $$
\noindent
for the differential rotation $\omega$ as a function of the specific 
external torque $f(\omega,r)$, Link, Epstein and Baym (1992) have argued that 
a frequency deficit {\it can not} be explained by the existing glitch models.
In other words, according to their work the postglitch frequency 
$\Omega_{c}(t)$ must be always greater than the extrapolated preglitch 
frequency $\Omega_{co}(t)$, contrary to the observations. On the other hand, 
Alpar \& Pines (1993) have discussed the theoretical interpretation of these 
events in the framework of vortex creep theory. Even though there seems to be 
enough room for such a behaviour in the latter, several details of the pinning 
layers are still unclear and complicate the interpretation. Thus, it may be 
interesting to explore alternatives to bring the theoretical picture closer to 
the observed phenomenology.

An important aspect of these observations is that, given the much larger 
amount of data taken from the Crab pulsar and the difficulties of extracting 
this signature from timing noise, it is entirely possible that other young 
pulsars also behave similarly. If so, events producing a permanent $\Delta 
{\dot \Omega}/ {\dot \Omega}$ may be important for the understanding of pulsar 
dynamics. Models having variable external torques have been considered in the 
past (see Blandford \& Romani 1988 and references therein) and may be helpful 
for understanding the data. Our goal in this work is to see to what extent a 
simple (but consistent) picture of young pulsar dynamics can be built by 
assuming a specific version of a variable external torque model. Section 2 is 
dedicated to formulate a minimal dynamical model in which the angle between 
the magnetic dipole and the rotation axis of the pulsar is allowed to vary 
according to simple laws. Section 3 presents an application of the model to 
the Crab, Vela and two other interesting pulsars (PSR B1509-58 and PSR B0540-69
). Finally, we present a discussion and conclusions in Section 4. There is also
an Appendix with exact solutions for the proposed growth laws.

\vskip 1 true cm
\noindent
{\bf 2 BASIC EQUATIONS}

The equation of motion of a rotating pulsar, assumed to have a crust+core 
normal component (suffix {\it c}) and a superfluid shell (suffix {\it s}) is 

$$ I_{c} \dot \Omega_{c} \; + \; I_{s} \dot 
\Omega_{s} \; = \; \tau_{ext} \eqno (1) $$
\noindent
since the frequency difference $\Omega_{c} - \Omega_{s} \; = \; \omega$ is 
constant on sufficiently long timescales, the rotational equilibrium implies 
further that 

$$ I_{tot} \dot \Omega_{c} \; = \; \tau_{ext}  . \eqno (2) $$
                         
Up to now studies have been based mainly on the use of torque laws of the 
form $\tau_{ext} \; = \; -K \Omega_{c}^{n}$, where $K$ is a constant (up to 
its possible decay for very old objects) and the braking index $n$ is 
assumed to be 3 as predicted by the magnetic dipole model (see e.g. 
Manchester \& Taylor 1977). The vacuum dipole model expression corresponds 
to $K \, = \, {2 \over {3 \, c^{3}}} \, \left\vert M \right\vert^{2} \, 
\sin^{2} \alpha$, where $\alpha$ is the angle between $M$ and $\Omega_{c}$ , 
$\left\vert M \right\vert \, = \, B_{o} \, R^{3}$ and $B_{o}$ , $R$ are 
the magnetic field and the radius of the star respectively.

From eq.(2) and the form of $\tau_{ext}$ it is clear that the peculiar 
events of 1975 and 1989 require either a reduction of $I_{tot}$, an increase 
of the magnitude of the magnetic moment $M$ or an increase of the 
angle $\alpha$ between $M$ and $\Omega_{c}$. Although all these hypothesis 
are in principle allowed by the data, we shall address here the change in the 
relative orientation of $M$ and $\Omega_{c}$ axis in those events. The idea is 
appealing not only because of its simplicity but also because it may be 
considered as a realisation of Ruderman's plate tectonic theory (Ruderman 1991)
where it finds a natural room. The other two hypothesis also have a theoretical
support in the framework of vortex creep theory (Alpar \& Pines 1993) and 
magnetic field generation or surfacing (Blandford, Applegate \& Hernquist 1883,
Muslimov \& Page 1996) respectively. They have been previously 
considered and we shall not address them here.  

At this point it must be noted that a counter-aligning pulsar in which $\alpha$ 
increases with time is quite remarkable, since the 
opposite situation is to be expected naively for the rotating/radiating star.
Therefore, the presence of "anomalous" glitch events is {\it per se} a 
signature of the complexity of the internal/magnetospheric dynamics which 
ultimately determine the behaviour of $\alpha (t)$. Several papers  have 
addressed the question of the internal dynamics and its consequences for 
$\alpha (t)$ (Michel \& Goldwire 1970, Davis \& Goldstein 1970, 
Goldreich 1970, Michel 1973, Lyne \& Manchester 1988, Michel 1991). Macy (1974)
has given a general treatment by solving simultaneously the system of equations
which determine $\alpha (t)$ and $\Omega_{c} (t)$ ($ \equiv \, \Omega_{pulse}$)
adopting a simple vacuum dipole model for the pulsar radiation. 
In order to check the consistency of the increasing angle hypothesis we 
have chosen instead to parametrise the growth of $\alpha$ by simple expressions 
and solve analytical models to compare them with the observations.

In first place, we want to know if the angle growth happens solely on 
glitches, in a discrete mode, or if there is a inter-glitch continuous
contribution. To do this, we can determine the variation of the angle $\alpha$
due to persistent shifts $\Delta {\dot \Omega} / {\dot \Omega}$ and $\Delta 
\Omega / \Omega$ which is easily found from eq.(2) to be

$$\Delta \alpha \, = \, {\biggl( {{\Delta \dot \Omega} \over {\dot \Omega}}  \,
 - \, 3{{\Delta \Omega} \over \Omega} \biggr)} \, {{\tan \alpha} \over 2}.
\eqno(3)$$

Dividing eq.(3) by $\Delta t$, here defined as a typical time-scale between 
glitches, we obtain a (mean) increase rate which we shall denote as $\big<
{{\Delta \alpha} \over {\Delta t}} \big>$. We shall keep in mind that, since
all we have at disposal is a very short observational span of $\sim 20 \, yr$
at most, there could be a discrepancy between $\big< {{\Delta \alpha} \over 
{\Delta t}} \big>$ and the continuous modeling in which $\dot \alpha$ is a
true local derivative. In fact, the data from Lyne, Pritchard \& Smith (1993) 
allows us to find for the Crab pulsar

$${\biggl< {{\Delta \alpha} \over {\Delta t}} {1 \over {\tan \alpha}} \biggr>} 
\, \simeq \, 1.8 \times 10^{-5} \, rad \, yr^{-1} \eqno(4)$$
\noindent
with $\Delta t \, = \, 4.6 \, yr$. On the other hand, the rate needed to 
account for the observed braking index is found from eq.(A14) to be

$${{\dot \alpha} \over {\tan \alpha}} \, = \, {{n_{obs} - 3} \over 2} {{\dot
\Omega} \over \Omega} \, = \, 9.6 \times 10^{-5} \, rad \, yr^{-1} \eqno(5)$$
\noindent
about 5 times the rate obtained from glitches. For PSR B0540-69, Vela and
PSR B1509-58, the rates given by eq.(5) are respectively $14.4 \times 10^{-5}
\, rad \, yr^{-1}$, $3.49 \times 10^{-5} \, rad \, yr^{-1}$ and $2.62 \times 
10^{-5} \, rad\, yr^{-1}$. All these values are within one order of magnitude, 
strengthening the idea that they could have the same origin. 
Once we have acknowledged that 
the main contribution to the angle growth comes of the interglitch 
slowdown, and because even pulsars (like PSR B1509-58 and PSR B0540-69) which 
did not display any glitches until now show low braking indexes, we have 
chosen to describe the angle growth as a {\it continuous} function rather than 
a discrete one. The other hypothesis put forward to explain the 
low braking index could be invoked here as above, but we will 
only explore the angle growth scenario.

We have tried four simple cases for the angle growth, namely an 
exponential, linear, power-law and logarithmic functions (see Appendix). 
Analytical expressions for $\Omega (t)$ have been obtained in all cases which, 
in principle, may be acceptable as descriptions of the long-term pulsar's 
spindown. However, a closer inspection of these $\Omega (t)$ forms reveals that
the linear and power-law solutions are not of general applicability because 
they critically depend on the specific input parameters (like the present angle
$\alpha_{p}$) to exist and be positive definite. Therefore, we have been led to 
consider the exponential growth $\alpha (t) \, = \, \alpha_{o} \, 
e^{t/t_{\alpha}}$ and the logarithmic growth $\alpha (t) \, = \, \ln  
{\bigl[ \, {t \over t_{p}} \, {\bigl( e^{\alpha_{p}} \, - \, e^{\alpha_{o}} 
\bigr)} \, + \, e^{\alpha_{o}} \, \bigr]}$ which do not suffer from these 
misbehaviours (see Appendix). 

We now turn to the application of a varying-angle model described by the 
formulae of the Appendix to specific cases in the next Section.

\vfill\eject
\vskip 1 true cm 
\noindent
{\bf 3 APPLICATION OF THE MODEL}

\bigskip
\noindent
{\it 3.1 General Considerations}

In order to check the consistency of the varying-angle model we shall 
apply it to the four pulsars in which $\ddot \Omega$ (and hence $n_{obs}$) 
has been measured. Starting from the Crab pulsar, in which a rather direct 
measure of $m_{obs}$ has been also 
obtained, we have attempted to unify as far as possible the behaviour of the 
young pulsar sample. The strategy is as follows: assuming that the braking 
index and jerk parameter discrepancies are solely due to the growth of 
$\alpha$, we impose the observed values of $n_{obs}$ and $m_{obs}$ 
(see eqs.(A17) and (A19) in the Appendix) to determine a solution for $\alpha$.
Once these values are obtained, eq.(A1), (A2) and (A3) can be used to 
calculate $P_{o}$, $\alpha_{o}$ and the time-scale for maximum torque (that 
is, the age at which $\alpha$ reaches $\pi/2$) termed $t_{m}$. 

\bigskip
\noindent
{\it 3.2 The Crab Pulsar (PSR B0531+21)} 

The Crab pulsar has been extensively studied and the basic quantities like 
$n_{obs}$ and $m_{obs}$ determined through measurements of $\Omega$ derivatives.
Using the observational data at MJD 40000.0 available from Lyne, Pritchard \& 
Smith (1993), we found for the exponential form (the logarithmic form does not 
provide a consistent solution for this pulsar)

$$\alpha_{p} \, \simeq \, 68^{o}$$

$$\alpha_{o} \, \simeq \, 56^{o}$$

$$P_{o} \, \simeq \, 19 \, ms \eqno(6)$$

$$t_{m} \, \simeq \, 2300 \, yr.$$

We note that the $t_{m}$ obtained is larger than the true age of the Crab 
$t_{p} \sim \, 940 \, yr$, as it should be. 
It is also worth mentioning that $\alpha_{p}$ is more consistent with the new
value inferred from optical polarization data ($\alpha_{p} \, \sim \, 60^{o}$, 
F.G.Smith et al. 1988) than the older one ($\alpha_{p} \, \geq \, 80^{o}$,
Kristian et al. 1970) and radio data (Rankin 1990).

As a corollary of the dynamical solution, the structural constant $C \, = \, 
{{2B_{o}^{2} R^{6}} \over{3c^{3} I_{tot}}}$ (see Appendix) can be now 
evaluated to be $4.14 \times 10^{-16} \, s$. If the magnetic field is taken 
to be $\sim 3.8 \times 10^{12} \,
G$ (Taylor, Manchester and Lyne 1993), a constraint on 
the product ${\bigl( {R \over {10^{6} \, cm}} \bigr)}^{6} \, 
{\bigl( {I \over {10^{45} \, g \, cm^{2}}} \bigr)}^{-1} \, 
\simeq \, 1.16 \, g \, cm^{-4}$ is obtained from the torque expression.
We thus see that the proposed dynamics offers some insight onto the 
equation of state, namely, the relationship favours a equation of state
slightly softer than the Bethe-Johnson I model (Bethe \& Johnson 1974).

\bigskip
\noindent
{\it 3.3 The Vela Pulsar (PSR B0833-45)} 

A measurement of $n_{obs}$ for the Vela pulsar has been recently obtained by 
Lyne et al. (1996), and it has not been possible determine $m_{obs}$ as yet, 
therefore the procedure used
above is not suitable. However, introducing the observational data from the
catalog of Taylor, Manchester \& Lyne (1993), we find that a logarithmic 
solution is possible only if $m_{obs} \, > \, 4.5$ (and 
\hbox{${\Omega}\!\!\!\!^{^{^{...}}}$} $> \, -8.6 \times 10^{-34} \, rad \, 
s^{-4}$), and an exponential one requires $m_{obs} \, < \, 3.2$ (and 
\hbox{${\Omega}\!\!\!\!^{^{^{...}}}$} $< \, -6.1 \times 10^{-34} \, rad \, 
s^{-4}$). An observational determination of the jerk parameter would thus
discriminate the angle-growth. Meanwhile, eq.(A15) provides an upper limit for 
$m_{obs}$ of 3.2, calculated {\it without} considering
the term which contains the angle and its derivatives. Therefore 
the only acceptable solution is a exponential growth of $\alpha$. For an 
arbitrary value of $\alpha_{p}$ we have found the limits 

$$\alpha_{o} \, < \, 32^{o}$$

$$P_{o} \, > \, 51 \, ms \eqno(7)$$

$$t_{m} \, < \, 1.4 \times 10^{5} \, yr$$
\noindent
where the carachteristic age has been used as the true age. There are claims
(Aschenbach, Egger \& Tr\"umper 1995, Lyne et al. 1996) suggesting that Vela is 
actually older by a factor of 2 or 3. 
This last possibility would imply in our analysis smaller limits for 
$\alpha_{o}$ and $P_{o}$ for the same $\alpha_{p}$, and a bigger limit for 
$t_{m}$ . An older pulsar also implies that $m_{obs}$ becomes a positive 
number. This is as far as we can go without any further data.

\bigskip
\noindent
{\it 3.4 PSR B0540-69} 

This pulsar does not have a observational determination of $m_{obs}$ either.
As an interesting feature of our models we predict using 
the data presented in Taylor et al. (1995) (Taylor, Manchester \& Lyne 1993)
a logarithmic angle growth if $m_{obs}
\, > \, 7.8$ (\hbox{${\Omega}\!\!\!\!^{^{^{...}}}$} $> \, -8.4 \times 10^{-31}
\, rad \, s^{-4}$), and an exponential one if $m_{obs} \, < \, 7.2$ 
(\hbox{${\Omega}\!\!\!\!^{^{^{...}}}$} $< \, -7.7 \times 10^{-31} \, rad \,
s^{-4}$). The upper limit from eq.(A15) is $m_{obs} \, \simeq \, 7.3$, so a
logarithmic growth is also excluded as in the case of Vela. Letting 
$\alpha_{p}$ to be a free parameter, we can calculate for an exponential growth

$$\alpha_{o} \, < \, 41^{o}$$

$$P_{o} \, > \, 23 \, ms. \eqno(8)$$

$$t_{m} \, < \, 3.3 \times 10^{4} \, yr$$

\bigskip
\noindent
{\it 3.5 PSR B1509-58} 

With the data by Kaspi et al. (1994) and using the characteristic age as $t_{p}
\, \simeq \, 1550 \, yr$, we find that 
the logarithmic growth is the only one allowed for this pulsar and gives

$$\alpha_{p} \, \simeq \, 83^{o}$$

$$\alpha_{o} \, \simeq \, 61^{o} \eqno(9)$$

$$P_{o} \, \simeq \, 45 \, ms$$

$$t_{m} \, \simeq \, 2200 \, yr.$$

Again, we can calculate the structural constant $C \, \simeq \, 6 \times
10^{-15} \, s$, which leads to ${\bigl( {R \over {10^{6} \, cm}} \bigr)}^{6} \, 
{\bigl( {I \over {10^{45} \, g \, cm^{2}}} \bigr)}^{-1} \, \simeq \, 1.01 \, g 
\, cm^{-4}$, assuming $B_{o} \simeq 1.55 \times 10^{13} \, G$ (Taylor, 
Manchester and Lyne 1993). This is 
approximately the same result obtained for the
Crab pulsar, or about 3\% relative difference in the radius of both objects, 
suggesting that the same equation of state can be applied to them. 

There is some dispute about this pulsar's true age (Thorsett 1992, Blandford
\& Romani 1988, Van den Bergh \& Kramper 1984). If our picture is correct,
the maximum admissible 
value for the pulsar's age $t_{p}$ is $\sim 1730 \, yr$ but in this case the
initial period would have been as low as $\sim 4 \, ms$, which seems difficult
to accept. These figures would support the
association of this pulsar with the 185 AD supernovae. On the other hand, if
this pulsar is older (the remnant is calculated to be $10^{4} \, yr$, see Van
den Bergh \& Kramper and references therein), its angle $\alpha$ may have 
reached $\pi /2$ in the past, so that our analisis no longer applies to it.
 
\vskip 1 true cm 
\noindent
{\bf 4 CONCLUSIONS}

We have studied simple parametric dynamical models of the angle growth 
which led us to determine the initial features of the four considered 
pulsars. Even if we have tried to unify as much as possible their 
dynamical behaviour, some ambiguity related to the intrinsic 
difficulty of the data analysis remains. As an example, 
the data analysis of the Crab 
pulsar made by Lyne, Pritchard \& Smith (1993) have not directly calculated  
the third derivative of frequency \hbox{${\nu}\!\!\!^{^{...}}$}, 
but instead proceeded to find it by setting $m_{obs} \, =$ 
\hbox{${\nu}\!\!\!^{^{...}}$} $\nu^{2} \, / \, {\dot \nu}^{3} \, = \,
n_{obs}(2n_{obs}-1)$. A inspection of eq.(A15) reveals that if the canonical
braking index is replaced by $n_{obs}$ (M.P.Allen \& J.E.Horvath, in
preparation) the angle-dependent term automatically vanishes, but since the
variation of $n_{obs}$ is small in the Crab, the correction is also small.
Strictly speaking, the true $m_{obs}$ in the case of a varying angle
would be greater than that value if the "angular" term

$${\Omega \over \dot \Omega} {\biggl( {\ddot \alpha \over \dot \alpha} - {{2 
\dot \alpha} \over {\sin ( 2 \alpha )}} \biggr)} \, < \, n_{obs} -1 \eqno(10)$$
\noindent
and smaller if the inequality is reversed.
If we substitute $\alpha_{p} \, = \, 68^{o}$ as found in Section 3.2 in 
eq.(A15), we obtain a correction to $m_{obs}$ of + 0.15. It 
should be noted that this correction in \hbox{${\Omega}\!\!\!\!^{^{^{...}}}$} 
implies small corrections in the other derivatives measurements
particularly in $\ddot \Omega$. Also, the discrepancy in the braking index, 
either originated by angle growth, magnetic field increase or inertia moment 
decrease, {\it necessarily} leads to some correction in the estimatives of 
$m_{obs}$ made through the simple assumption above, unless fortuitous
cancellations occur.

We consider that an exponential angle growth, with a typical (but 
non-universal) time-scale $\leq 10^{4} \, yr$ is a good (yet simple) model for 
the young pulsars dynamics. It accounts for the observed discrepancies in the 
braking index and the jerk parameter. It also provides limits for the initial 
parameters of individual pulsars if just $n_{obs}$ is known. These preditions 
can be compared to the observational data to be gathered in a few years.

As a final remark we stress that our results show that the characteristic age
may be a rather poor estimative of the true age. The dependence of the torque
on $t$ forces a full calculation of $t_{p}$ by inverting the complicated
expression of eq.(A1) instead of using the standard form

$$\tau \, = \, - \, {\Omega_{p} \over {(n-1) \dot \Omega_{p}}} {\biggl[ 1 \, - 
\, {\biggl( {\Omega_{p} \over \Omega_{o}} \biggr)}^{n-1} \biggr]}, \eqno(11)$$
\noindent
which nevertheless remains valid (even if $\dot \alpha \neq 0$) in $n_{obs} \, =
\, constant$ (M.P.Allen \& J.E.Horvath, in preparation). 
We conclude that, even if a exponential growth 
of the $\alpha$ angle seems to be consistent with several observed aspects of 
young pulsars behaviour, more observational data (mainly a detection of     
a change in the average pulse profile at the level $\sim \, 10^{-3}$) is      
needed to formulate more complete and complex models of the dynamics. 
Reliable determinations of $\alpha$ for these pulsars through optical
polarization or other methods and $m_{obs}$ would be decisive to test
angle-growing models.

While this paper was being written we have received a preprint by B.Link \&
R.I.Epstein where some of these ideas were independently exposed and discussed.
\vskip 1 true cm
\noindent
{\bf 5 ACKNOWLEDGEMENTS}

We would like to acknowledge S.O.Kepler and C.Beaug\'e for encouragement 
and advice during this work. We are grateful to F.Camilo which has provided us
details of his joint work in advance and provided guidance to the catalog data.
The useful advice of anonymous referees contributed to inprove a former version
of this work. This work has been partially supported by the CNPq (Brazil) 
through a Research Fellowship (J.E.H.) and CAPES (Brazil) to M.P.A..

\vfill\eject 
\noindent
{\bf 6 REFERENCES}

\bigskip
\noindent
Alpar,M.A., Anderson,P.W., Pines,D., Shaham,J., 1984, ApJ, 276, 325
\bigskip
\noindent
Alpar,M.A., Pines,D., 1993, in Isolated Pulsars (eds. van
Riper,K.A., Epstein,R.I., Ho,C.), 17 (Cambridge University Press)
\bigskip
\noindent
Aschembach,B., Egger,R., Kramper,K.W., 1995, Nat, 373, 587
\bigskip
\noindent
Baym,G., Pethick,C., Pines,D., Ruderman,M., 1969, Nat, 224, 872
\bigskip
\noindent
Bethe,H., Johnson,M., 1974, Nuc.Phys.A, 230, 1
\bigskip
\noindent
Davis,L., Goldstein,M., 1970, ApJ, 159, L81
\bigskip
\noindent
Blandford,R.D., Applegate,J.H., Hernquist,L., 1983, MNRAS, 204, 1025
\bigskip
\noindent 
Blandford,R.D., Romani,R.W., 1988, MNRAS, 234, 57
\bigskip
\noindent
Demia\'nski,M., Pr\'oszy\'nski,M., 1983, MNRAS, 202, 437
\bigskip
\noindent
Goldreich,P., 1970, ApJ, 160, L11
\bigskip
\noindent
Gullahorn,G.E. et al., 1977, AJ, 82, 309
\bigskip
\noindent
Jones,P.B., 1990, MNRAS, 243, 257
\bigskip
\noindent
Kaspi,V. et al., 1994, ApJ, 422, L83
\bigskip
\noindent
Kristian,J. et al., 1970, ApJ, 162, 475
\bigskip
\noindent
Link,B., Epstein,R., Baym,G., 1992, ApJ, 390, L21
\bigskip
\noindent
Link,B., Epstein,R., Baym,G., 1993, ApJ, 403, 285
\bigskip
\noindent
Lohsen,E.H.G., 1981, A\&A, 44, 1
\bigskip
\noindent
Lyne,A.G., Manchester,R.N., 1988, MNRAS, 234, 477
\bigskip
\noindent
Lyne,A.G., Pritchard,R.S., 1987, MNRAS, 229, 223
\bigskip
\noindent
Lyne,A.G., Pritchard,R.S., Smith,F.G., Camilo,F., 1996, Nat, 381, 497
\bigskip
\noindent
Lyne,A.G., Smith,F.G., Pritchard,R.S., 1992, Nat, 359, 406
\bigskip
\noindent
Lyne,A.G., Smith,F.G., Pritchard,R.S., 1993, MNRAS, 265, 1003
\bigskip
\noindent
Macy,W.W., 1974, ApJ, 190, 153
\bigskip
\noindent
Manchester,R.N., Taylor,J.H., 1977, Pulsars (Freeman, San Francisco)
\bigskip
\noindent
Michel,F.C., 1973, ApJ, 180, L133
\bigskip
\noindent
Michel,F.C., 1991, Theory of Pulsar Magnetospheres (University of Chicago
Press, Chicago)
\bigskip
\noindent
Michel,F.C., Goldwire,H.C., 1970, ApLett, 5, 21
\bigskip
\noindent
Muslimov,A., Page,D.N., 1996, ApJ, 458, 347
\bigskip
\noindent
Pines,D., Alpar,M.A., 1985, Nat, 316, 27
\bigskip
\noindent
Rankin,J.M., 1990, ApJ, 352, 247
\bigskip
\noindent
Ruderman,M., 1991, ApJ, 366, 261
\bigskip
\noindent
Smith,F.G., Jones,D.H.P., Dick,J.S.B., Pike,C.D., 1988, MNRAS, 233, 305
\bigskip
\noindent
Taylor,J.H., Manchester,R.N., Lyne,A.G., 1993, ApJ Sup. Ser., 88, 529
\bigskip
\noindent
Taylor,J.H., Manchester,R.N., Lyne,A.G., Camilo,F., 1995, Unpublished work
\bigskip
\noindent
Thorsett,S.E., 1992, Nat, 356, 690
\bigskip
\noindent
Van den Bergh,S., Kramper,K.W., 1984, ApJ, 280, L51

\vfill\eject
\vskip 1 true cm 
\noindent
{\bf 7 APPENDIX}

We present in this Appendix the explicit forms $\alpha(t)$ employed for the
determination of the dynamical pulsar histories. First let us consider an 
exponential growth of the angle $\alpha (t) \, = 
\, \alpha_{o} e^{t/t_{\alpha}}$. Assuming a vacuum dipole model for 
$\tau_{ext}$, eq.(2) can be integrated to yield the analytic form for $\Omega$ 

$$ \Omega \, = \, \Omega_{p} {\biggl[ \, 1 + {t_{\alpha} \over {T \sin^{2} 
\alpha_{p}}} {\sum_{n=1}^{\infty}} \, (-1)^{n} \, {{(2 \alpha_{p})^{2n} \, - \,
(2 \alpha_{o})^{2n}} \over {(2n) \, (2n)!}} \, \biggr]}^{-1/2}  \eqno(A1) $$
\noindent
where $T \, = \, - \, {\Omega_{p} \over \dot\Omega_{p}} \, = \, {\bigl( C 
\sin^{2} \alpha_{p} \Omega_{p}^{2} \bigr)}^{-1}$ is a slowdown time-scale at the
present time, with $C$ being a structural constant depending on the chosen 
equation of state, $\alpha_{p}$ and $\alpha_{o}$ are the present and initial 
angle respectively, and $t_{\alpha} \, = \, \alpha_{p} / \dot \alpha_{p}$ is a 
constant time-scale for the growth of the angle. Now, we can express the
time-scale for maximun torque $t_{m}$, the initial period $P_{o}$ and 
$\alpha_{o}$ in terms of observable quantities yielding            
                                                                             
$$ t_{m} \, = \, t_{\alpha} \ln {\biggl( {\pi \over {2 \alpha_{p}}} \biggr)} + 
t_{p} \eqno(A2)$$                            
      
$$ \alpha_{o} \, = \, \alpha_{p} e^{-t_{p}/t_{\alpha}}. \eqno(A3)$$

If we assume instead a linear growth law  $\alpha (t)
\, = \, {\bigl( {\pi \over {2}} - \alpha_{o} \bigr)} {t \over {t_{m}}} \, + \, 
\alpha_{o}$ and integrate again the equation of motion, we obtain the 
non-linear algebraic equation               
                                                                             
$$\Omega \, = \, \Omega_{p} {\biggl[ \, 1 - {1 \over {T \dot \alpha \sin^{2} 
\alpha_{p}}} \, {\biggl( \alpha_{p} \, - \, \alpha \, - \, \sin (2 \alpha_{p}) 
\, + \, \sin (2 \alpha) \biggr)} \, \biggr]}^{-1/2} \eqno(A4)$$              
\noindent                                                    
which gives                                               
                                                                             
$$ t_{m} \, = \, {\biggl( {\pi \over 2} \, - \, \alpha_{p} \biggr)} \,
{1 \over \dot \alpha} \, + \, t_{p} \eqno(A5)$$     
                                               
$$ \alpha_{o} \, = \, \alpha_{p} \, - \, \dot \alpha \, t_{p}. \eqno(A6)$$
 
Integration of a logarithmic law $\alpha (t) \, = \, \ln {\bigl[ \, {t \over
t_{p}} \, {\bigl( e^{\alpha_{p}} \, - \, e^{\alpha_{o}} \bigr)} \, + \, 
e^{\alpha_{o}} \, \bigr]}$ gives 

$$ \Omega \, = \, \Omega_{p} \biggl\{ \, 1 - {{2 \, t_{p}} \over {5T {\bigl( 
e^{\alpha_{p}} - e^{\alpha_{o}} \bigr)} \sin^{2} \alpha_{p} }} \, \times$$

$$\times \, {\biggl[ e^{\alpha_{p}} \, {\biggl( \sin^{2} \alpha_{p} - \sin (2 
\alpha_{p}) + 2 \biggr)} - e^{\alpha} \, {\biggl( \sin^{2} \alpha - \sin (2 
\alpha) + 2 \biggr)} \biggr]} \biggr\}^{-1/2} \eqno(A7)$$ 
\noindent
with the following results

$$ t_{m} \, = \, {{e^{\pi /2} \, - \, \alpha_{p} \, - \, \ln {\bigl( 1-t_{p}
\dot \alpha_{p} \bigr)}} \over {e^{\alpha_{p}}} \, \dot \alpha_{p}} \eqno(A8)$$

$$ \alpha_{o} \, = \, \alpha_{p} \, + \, \ln {\bigl( 1 \, - \, t_{p} \, \dot
\alpha_{p} \bigr)}. \eqno(A9)$$

Finally, integrating a power-law $\alpha (t) \, = \, {\bigl( {\pi \over 2} -
\alpha_{o} \bigr)} {\bigl( {t \over t_{m}} \bigr)}^{b} \, + \, \alpha_{o}$ ,
where $b$ is a positive constant $\neq$ 1, we have

$$ \Omega \, = \, \Omega_{p} \biggl\{ \, 1 - {{t_{p} \, t_{m}} \over {T
{\bigl( {\pi \over 2} - \alpha_{o} \bigr)}^{1/b} \sin^{2} \alpha_{p}}} \,
\biggl[ ( \alpha_{p} - \alpha_{o} )^{1/b} - {\bigl( \alpha - 
\alpha_{o} \bigr)}^{1/b} - {\biggl( {1 \over b} \biggr)} ! \, \times $$ 

$$ \times \, \biggl[ \cos (2 \alpha_{p} ) {\sum_{i=0}^{int( {1 \over 2b} -1)}} 
f_{2i+2} (\alpha_{p}) \, - \, \cos (2 \alpha ) {\sum_{i=0}^{int 
( {1 \over 2b} -1)}} f_{2i+2} (\alpha) \, +$$

$$+ \, \sin (2 \alpha_{p} ) {\sum_{i=0}^{int( {1 \over 2b} - {1 \over 2} )}} 
f_{2i+1} (\alpha_{p}) \, - \, \sin (2 \alpha )
{\sum_{i=0}^{int( {1 \over 2b} - {1 \over 2} )}} f_{2i+1} (\alpha) \biggr] 
\biggr] \biggr\}^{-1/2} \eqno(A10)$$
\noindent   
where  

$$f_{2i+1} (\alpha) \, = \, (-1)^{i} {{(\alpha - \alpha_{o} )^{{1 \over b} 
-2i-1}} \over {2^{2i+1} {\bigl( {1 \over b} -2i-1 \bigr)} !}}$$

$$f_{2i+2} (\alpha) \, = \, (-1)^{i} {{(\alpha - \alpha_{o} )^{{1 \over b} 
-2i-2}} \over {2^{2i+2} {\bigl( {1 \over b} -2i-2 \bigr)} !}} \, ,$$

and $int(x)$ is the greatest integer number lesser than $x$, for $b= {\bigl\{
{1 \over 2} , {1 \over 3} , {1 \over 4} , ... \bigr\}}$. For any other $b$ we 
will have

$$ \Omega \, = \, \Omega_{p} \biggl\{ \, 1 - {{t_{p} \, t_{m}} \over {T
{\bigl( {\pi \over 2} - \alpha_{o} \bigr)}^{1/b} \sin^{2} \alpha_{p}}} \,
\biggl[ ( \alpha_{p} - \alpha_{o} )^{1/b} {\bigl( \cos^{2} \alpha_{o} - \cos 2
\alpha_{p} \bigr)} \, -$$

$$- \, {\bigl( \alpha - \alpha_{o} \bigr)}^{1/b} {\bigl( \cos^{2}
\alpha_{o} - \cos 2 \alpha \bigr)} \, - \, {\biggl( {1 \over b} \biggr)} ! 
\biggl[ \cos (2 \alpha_{p} ) {\sum_{i=1}^{\infty}} g_{2i} (\alpha_{p}) \, - \, 
\cos (2 \alpha ) {\sum_{i=1}^{\infty}} g_{2i} (\alpha) \, +$$

$$+ \, \sin (2 \alpha_{p} ) {\sum_{i=0}^{\infty}} g_{2i+1} (\alpha_{p}) \, - \,
\sin (2 \alpha ) {\sum_{i=0}^{\infty}} g_{2i+1} (\alpha) \biggr] \biggr] 
\biggr\}^{-1/2} \eqno(A11)$$
\noindent   
where

$$g_{2i} (\alpha) \, = \, (-1)^{i} {{2^{2i} (\alpha - \alpha_{o} )^
{{1 \over b} +2i}} \over {{\bigl( {1 \over b} +2i \bigr)} !}}$$

$$g_{2i+1} (\alpha) \, = \, (-1)^{i} {{2^{2i+1} ( \alpha - \alpha_{o} )^{{1 
\over b} +2i+1}} \over {{\bigl( {1 \over b} +2i+1 \bigr)} !}}$$
\noindent
giving

$$t_{m} \, = \, t_{p}^{(1- {1 \over b})} \, {\biggl[ {{{\bigl( {\pi \over 2} - 
\alpha_{p} \bigr)} b} \over {\dot \alpha_{p}}} \biggr]}^{1/b} \eqno(A12)$$

$$\alpha_{o} \, = \, \alpha_{p} \, - \, {{t_{p} \dot \alpha_{p}} \over b}.
\eqno(A13)$$

Note that the expressions eqs.(A1), (A4), (A7), (A10) and (A11) must be now used 
to define the characteristic age of the pulsar replacing the usual form $\tau 
\, = \, {1 \over {(n \, - \,1)}} \, {\Omega \over {\left\vert \dot\Omega 
\right\vert}}$ (see Section 4).

It also follows that in these angle varying models it can be shown that, even 
if the power of $\Omega$ in the torque expression is exactly $3$, the observed 
braking index defined as $n_{obs} \, = \, \ddot \Omega \, \Omega / \dot 
\Omega^{2}$ picks up an extra term 

$$ n_{obs} \, = \, 3 \, - \, 2 \, \left\vert {\Omega \over {\dot \Omega}} 
\right\vert  \, \cot \alpha \, {d\alpha \over {dt}} \eqno(A14) $$
\noindent
and the observed jerk parameter defined as $m_{obs} \, =$ 
\hbox{${\Omega}\!\!\!\!^{^{^{...}}}$}$\, \Omega^{2} / \dot \Omega^{3}$ is
also modified

$$ m_{obs} \, = \, 3 \, (2 \, n_{obs} -1)+(n_{obs} -3) {\biggl[ n_{obs} +
{\Omega \over \dot \Omega} {\biggl( {\ddot \alpha \over \dot \alpha} - {{2 \dot
\alpha} \over {\sin ( 2 \alpha )}} \biggr)} \biggr]}. \eqno(A15)$$

Therefore, using the simple expressions for $\alpha (t)$ we obtain

$${{\sin (2 \alpha )-2 \alpha} \over {4 \cos^{2} \alpha}} \, = \, {{m_{obs} 
-n_{obs} (n_{obs} +3) +3} \over {(n_{obs} -3)^{2}}} \eqno(A16)$$
\noindent   
for the exponential form, 

$$\cos \alpha \, = \, {\biggl[ -2 \, {\biggl( {{m_{obs} -n_{obs} (n_{obs} +3) 
+3} \over {(n_{obs} -3)^{2}}} \biggr)} \biggl]}^{-1/2} \eqno(A17)$$
\noindent
for the linear form,

$${{\sin (2 \alpha )+2} \over {4 \cos^{2} \alpha}} \, = \, {{m_{obs} 
-n_{obs} (n_{obs} +3) +3} \over {(n_{obs} -3)^{2}}} \eqno(A18)$$
\noindent
for the logarithmic form, and

$$\cos \alpha \, = \, {\biggl[ -2 \, {\biggl( {{m_{obs} -n_{obs} (n_{obs} +3) 
+3} \over {(n_{obs} -3)^{2}}} + {{T \, (b-1)} \over {t_{p} \, 
(n_{obs} -3)}} \biggr)} \biggr]}^{-1/2} \eqno(A19)$$
\noindent
for the power-law. These forms express the unknown angles in terms of the
$n_{obs}$ and $m_{obs}$.

\bye